\documentclass[letterpaper]{article} 
\usepackage{aaai23}  
\usepackage{times}  
\usepackage{helvet}  
\usepackage{courier}  
\usepackage[hyphens]{url}  
\usepackage{graphicx} 
\def\UrlFont{\rm}  
\usepackage{natbib}  
\usepackage{caption} 
\usepackage{algorithm}
\usepackage{algorithmic}

\usepackage{amsfonts}
\usepackage{amsmath}
\usepackage{amsthm}
\usepackage{mathtools}
\usepackage{epsfig}
\usepackage{subfigure}
\usepackage{array}
\usepackage{lipsum}
\usepackage{bm}
\usepackage{textcomp}
\usepackage{enumitem}
\usepackage{multirow}
\usepackage{epstopdf}

\newtheorem{definition}{Definition}

\usepackage{newfloat}
\usepackage{listings}
\floatstyle{ruled}
\newfloat{listing}{tb}{lst}{}
\floatname{listing}{Listing}
\title{InfluencerRank: Discovering Effective Influencers via Graph Convolutional Attentive Recurrent Neural Networks}
\author {
    Seungbae Kim\textsuperscript{\rm 1},
    Jyun-Yu Jiang\textsuperscript{\rm 2},
    Jinyoung Han\textsuperscript{\rm 3} and
    Wei Wang\textsuperscript{\rm 2} \\
}
\affiliations {
    \textsuperscript{\rm 1} Department of Computer Science and Engineering, University of South Florida \\
    \textsuperscript{\rm 2} Department of Computer Science, University of California, Los Angeles \\
    \textsuperscript{\rm 3} Department of Applied Artificial Intelligence, Sungkyunkwan University \\
    seungbae@usf.edu, xchen@cs.ucla.edu, jyunyu@cs.ucla.edu, jinyounghan@skku.edu, weiwang@cs.ucla.edu
}

\begin{document}

\maketitle

\begin{abstract}
As influencers play considerable roles in social media marketing, companies increase the budget for influencer marketing. Hiring effective influencers is crucial in social influencer marketing, but it is challenging to find the right influencers among hundreds of millions of social media users. In this paper, we propose \emph{InfluencerRank} that ranks influencers by their effectiveness based on their posting behaviors and social relations over time.
To represent the posting behaviors and social relations, the graph convolutional neural networks are applied to model influencers with heterogeneous networks during different historical periods.
By learning the network structure with the embedded node features, \emph{InfluencerRank} can derive informative representations for influencers at each period.
An attentive recurrent neural network finally distinguishes highly effective influencers from other influencers by capturing the knowledge of the dynamics of influencer representations over time.
Extensive experiments have been conducted on an Instagram dataset that consists of 18,397 influencers with their 2,952,075 posts published within 12 months. 
The experimental results demonstrate that \emph{InfluencerRank} outperforms existing baseline methods. An in-depth analysis further reveals that all of our proposed features and model components are beneficial to discover effective influencers.
\end{abstract}

\section{Introduction}\label{sec:introduction}

Influencers are known as individuals who influence a magnificent number of people on social media.
This, in turn, has attracted great attention to marketers since influencers and their huge fan bases can be considered as marketing channels and audiences, respectively~\cite{de2017marketing, evans2017disclosing}. More recently, companies have started hiring influencers to advertise products for targeted audiences and expand brand awareness. 

Due to the rapid growth of social media and influencer marketing, discovering effective influencers on social media has become increasingly important~\cite{riquelme2016measuring,kang2018understanding,kim2020profile}.
For measuring user influence on social media, well-known metrics, such as the numbers of followers, retweets, and mentions, have been widely applied~\cite{bakshy2011everyone, segev2018measuring}. In addition, information propagation~\cite{romero2011influence, silva2013profilerank, kempe2003maximizing}, social connections~\cite{li2011discovering}, network centrality~\cite{chen2017interplay}, transparency~\cite{kim2021discovering}, and multi-relational network~\cite{ma2018influencer} have been used to identify influencers on social media.
Among the various measures, the \emph{effectiveness of influence}~\cite{liu2015identifying}, often measured by the engagement rate~\cite{de2017marketing, kim2020detecting, kim2021evaluating, lou2019influencer, william2018engagement}, has been considered as crucial in identifying effective influencers especially in the marketing domain. The engagement rate can be calculated as the ratio of the average number of likes to the number of followers, which essentially shows how much audiences engage with the corresponding influencer.

\begin{figure}[!t]
\centering
\includegraphics[width=0.95\linewidth]{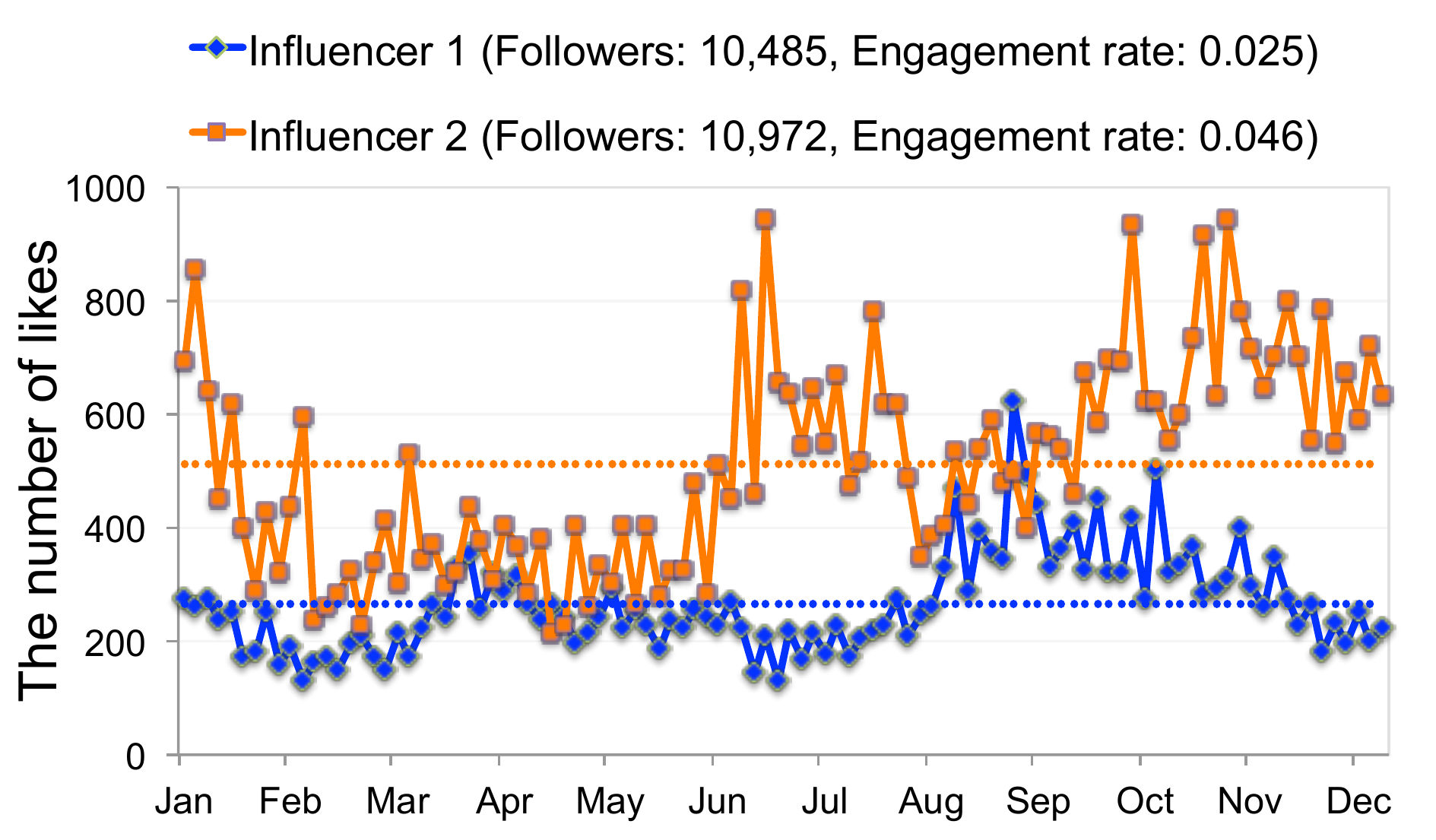}
\caption{The number of likes on posts published by two influencers across the time. Although two influencers have similar numbers of followers, the average numbers of likes (i.e., dotted lines) are significantly different. Additionally, the number of likes dynamically changes over time.}
\label{fig:case_study}
\end{figure}

To discover the effective influencers (i.e., influencers with high engagement rates), previous work used posting behaviors of influencers or characteristics of their posts.
For example, \citet{romero2011influence}, \citet{liu2015identifying}, and \citet{feng2018inf2vec} utilized the social networks among influencers; 
\citet{li2011discovering} analyzed post contents to derive statistical features in identifying influencers.
However, none of these studies jointly and comprehensively modeled posting behaviors, post characteristics, and social networking behaviors, which may result in a biased or partial representation of the effectiveness of influencer. 
For example, as shown in Figure~\ref{fig:case_study}, two influencers have similar numbers of followers hence they may be considered as having similar effectiveness, but their actual engagement rates are shown to be significantly different.
Although some methods applied the PageRank algorithm on influencer-content graphs~\cite{silva2013profilerank} and independently derived the features of influencers and posts~\cite{liu2015identifying}, the PageRank algorithm can be biased to a certain type of nodes~\cite{brezinski2006pagerank} when the relations between influencers and posts are ignored by independent features.

To address this issue, we propose to use a heterogeneous network to model the effectiveness of influencers with their posting behavior, social networking behavior, and post characteristics together.
In addition, considering historical behavioral patterns can be further beneficial to discover effective influencers since posting behavior of an influencer can change dynamically over time.
For instance, as shown in Figure~\ref{fig:case_study}, although an influencer does not receive many likes in the most recent time period, he/she may receive many number of likes in the future if he/she was used to get great attention in the past.
Moreover, analyzing time-varying behavior patterns can provide more evidence on the robustness of an influencer.
For example, the unstable performance (or effectiveness) of an influencer over time may not be desirable even if he/she satisfies the performance in the most recent time period.
Hence, taking such time-varying behavior patterns into account for discovering effective influencers is essential.
However, most of the prior studies only focused on the most recent information without considering the historical patterns of influencers.

In this paper, we propose \textit{InfluencerRank}, a learning framework, that discovers effective influencers in social media by learning historical behavioral patterns of influencers.
For comprehensively representing the effectiveness of an influencer, we build a heterogeneous information network that consists of influencers, hashtags, user tags, and image objects used by influencers for each historical time period~\cite{yang2020heterogeneous,zhang2019heterogeneous}.
To learn the complex posting behaviors, social networking, and post characteristics of each influencer, we apply graph convolutional networks (GCNs)~\cite{kipf2016semi} with well-designed influencer features, thereby deriving the influencer representation at a certain period. 
Based on the influencer representations over different historical time periods, the attentive recurrent neural network is proposed to learn the sequential and temporal behaviors to derive an ultimate representation.
Finally, a learning-to-rank framework ranks a list of influencers to discover the ones who are more effective than others.

We summarize our contributions as follows:
\begin{itemize}
\item To the best of our knowledge, this is the first attempt to rank influencers with their effectiveness by learning their historical behavioral patterns in social media marketing. We believe our model can be used for influencer recommendations that help companies to recruit a set of effective influencers to boost the advertising effect in social media. 
\item The \textit{InfluencerRank} uses the graph convolutional networks over general social media features to learn the posting behavior of the influencers as well as the characteristics of their posts, thus it can be applied to any social media to discover effective influencers.
Besides, recurrent neural networks are also applied to model sequential and historical behaviors of influencers over time.
We conduct experiments on a real-world dataset collected from Instagram~\cite{kim2020profile}, one of the most popular social media for influencer marketing~\cite{nanji2017instagram}. The results demonstrate that \textit{InfluencerRank} outperforms other existing methods for identifying effective influencers.
\item Our analysis further reveals that the image object nodes have more impact on discovering effective influencers than other types of nodes since they can densely connect influencers thereby removing noises in the network. We also find that user reactions and visual perception of images are important features to find effective influencers.
\item We evaluate our proposed model over groups of influencers with different numbers of followers, and highlight that \textit{InfluencerRank} shows effective and robust ranking performances across various groups of influencers.
\end{itemize}

\section{Related Work}\label{sec:related}

\subsection{Influence Prediction in Social Networks}

To find influencers in social networks, most studies rely on social media features to measure the influence. For example, the number of followers, posts, reposts, and mentions are well-known metrics to measure the influence of a user~\cite{bakshy2011everyone,subbian2011supervised}.
Based on the measures, \citet{bakshy2011everyone} use the regression tree model and \citet{subbian2011supervised} aggregate rank results to rank influencers, respectively. 
\citet{romero2011influence} propose the passivity of nodes to measure how likely the information is propagated in the social networks, and then apply the PageRank to rank the users.
\citet{liu2015identifying} consider the time domain over the user trust network in the proposed framework to classify influencers into one of three categories, emerging influencers, holding influencers, and vanishing influencers.
In addition to social network features, some studies propose to use machine learning with statistical features. \citet{li2011discovering} extract network-based, content-based, and user activeness-based statistical features, e.g., the number of followers, and length of posts, to predict the influence of users.
\citet{segev2018measuring} use simple statistics of posts and users, e.g., the number of likes, comments, followers, and posts, to measure the user influence using a regression model.
Some previous works, on the other hand, exploit graphical information.
\citet{zhang2015influenced} exploit the social influence locality to predict retweet behaviors.
\citet{qiu2018deepinf} utilize mini-batches of sub-graphs and apply the attention mechanism to predict the influence of users on social networks.
\citet{chen2019information} propose recurrent convolutional networks to consider temporal effect on information cascade prediction.
However, most previous works fail to consider temporal dynamics in the social relationships and characteristics of users.

\subsection{Graph Convolutional Recurrent Networks}
Graph convolutional networks~(GCNs)~\cite{kipf2016semi} are the neural network architecture for graph-structured data.
GCNs deploy spectral convolutional structures with localized first-order approximations so that the knowledge of both node features and graph structures can be leveraged.
However, while real-world data that can be modeled as graphs dynamically changes over time, temporal information cannot be easily captured from GCNs.
To learn the temporal dynamics of structural graphs, previous studies suggest to combine GCN and recurrent neural networks~(RNNs).
\citet{seo2018structured} propose the models that (i) stack up graphs to make RNN inputs and (ii) consider convolutions in RNNs, which can learn a sequence of structural information.
They find that each model outperforms the other models depending on applications such as video prediction and natural language modeling.
\citet{pareja2020evolvegcn} propose another approach to capture graph dynamics.
Instead of using a sequence of graph embedding as inputs of RNN, they first use RNN to acquire the knowledge of network parameter dynamics.
This approach can benefit in a case where a node dynamically appears and disappears. 
This paper proposes to apply attentive recurrent networks over the temporal node representations by taking a heterogeneous network that consists of influencers, image objects, hashtags, and user tags over time.
Our model can effectively learn temporal graph representations by estimating the importance of hidden states for certain time periods.

\section{Problem Statement}\label{sec:problem}

In this section, we formally define the effectiveness metric of an influencer and then formulate the problem of discovering effective influencers.
\begin{definition}
  \textbf{Engagement rate} is a widely-used metric in influencer marketing that shows how much audiences actively engage with an influencer~\cite{de2017marketing, kim2021evaluating, kim2020detecting, lou2019influencer, william2018engagement}. Given an influencer $u$, the engagement rate of the influencer at time $t$ is calculated as follows: 
  \begin{equation} \label{eq_1}
  E_{u}^{t}=\frac{l_{u}^{t}}{f_u}
  \end{equation}
  where $f_u$ is the number of followers who follow the influencer $u$ and $l_{u}^{t}$ is the average number of likes on content posted by the influencer $u$ at timestamp $t$. 
\end{definition}
Based on the definition of influencer effectiveness, we introduce the influencer ranking problem.
Let $U$ be the set of influencers.
For each timestamp $t$, we suppose that an influencer $u$ has published a set of posts $P^t_u$.
Given the set of influencers~$U$ and their posts published until time $k$, $\lbrace P^t_u \mid 1 \leq t \leq k \rbrace$, the goal of this work is to discover influencers with high engagement rates at time $k$ by ranking all influencers $u\in U$ so that $E_{u_i}^k$ is greater than $E_{u_j}^k$ if the influencer $u_i$ is ranked higher than the influencer $u_j$.


\section{Influencer Ranking Model Framework}\label{sec:model}

\begin{figure*}[!t]
    \centering
    \includegraphics[width=0.9\linewidth]{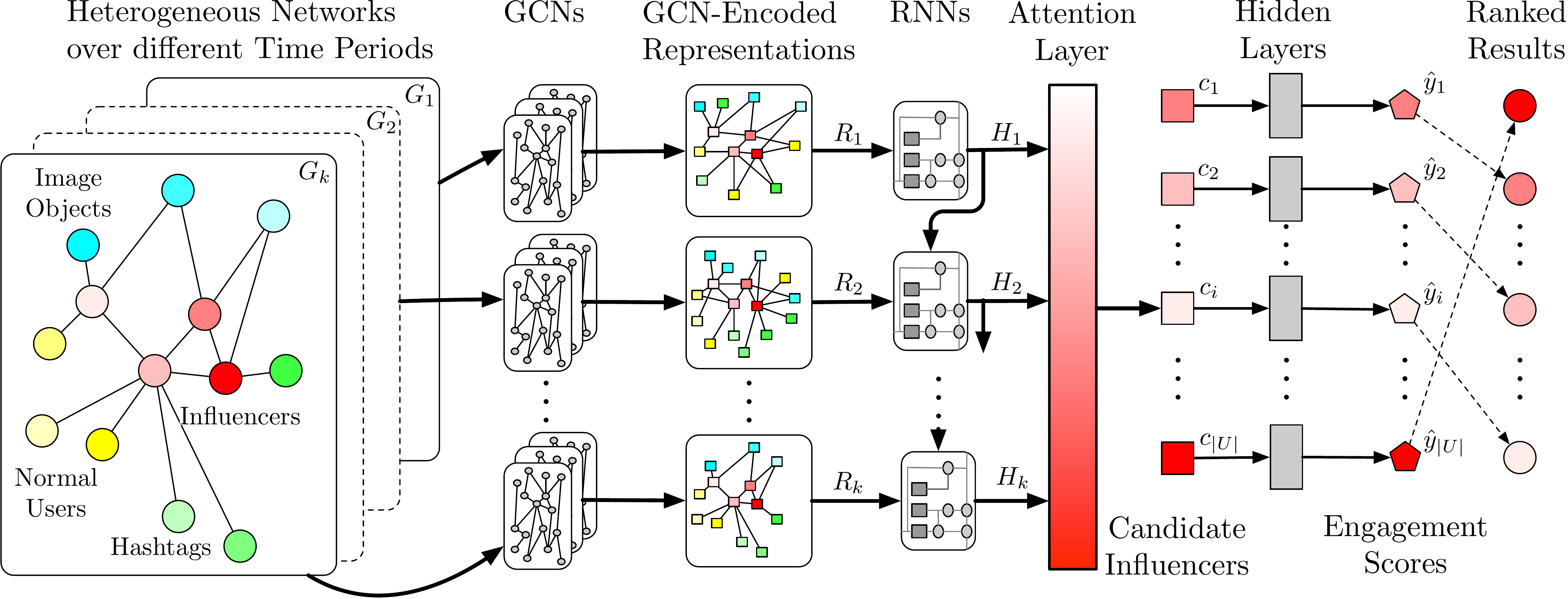}
    \caption{The overall framework of the proposed \textit{InfluencerRank}.}
    \label{fig:framework}
\end{figure*}

In this section, we propose InfluencerRank that learns the temporal dynamics of the engagement rates of influencers to automatically discover highly effective influencers.
Figure~\ref{fig:framework} shows the overall framework of the proposed InfluencerRank.
The framework takes a series of influencer social networks as input, where each network is composed of influencers and different entities, including but not limited to image objects, hashtags, and other users in social media.
The graph convolutional networks~(GCNs) are then applied to the input social networks to derive appropriate node representations that capture social relationships and posting characteristics of influencers at a certain time.
The GCN-encoded representations across different times are then fed into a recurrent neural network to learn from the sequence of the node representations.
The attention mechanism is then applied to the whole sequence of representations to finally derive the effectiveness scores of candidate influencers and rank them for discovering effective influencers.

\begin{table}[!t]
    \centering
    \resizebox{\linewidth}{!}{\begin{tabular}{|c|p{.8\linewidth}|} \hline
    Notation & Description  \\ \hline \hline
      $E^t_u$   & the engagement rate of an influencer $u$ at time $t$. \\
      $l^t_u$   & the average number of engagements on contents posted by the influencer $u$ at time $t$.\\
      $f^t_u$   & the number of followers for an influencer $u$ at time $t$.  \\
         \hline
      $U$ & the set of influencers. \\
      $P^t_u$ & the posts published by the influencer $u$ at time $t$.  \\ 
      $G_t$ & the heterogeneous network for time $t$ with the node features $X_t$ and the adjacency matrix $A_t$. \\ \hline
      $\hat{\bm{A_t}}$ & Normalized adjacency matrix transformed from $\bm{A_t}$. \\
      $d$ & the number of dimensions for embedded node features.\\
      $\bm{D}$ & the diagonal degree matrix of $\bm{A_t}$.\\
      $r$ & the number of hidden dimensions in GCNs.\\
      $\bm{F^{(i)}}$ & the outputs of the $i$-th GCN layer.\\
      $\bm{W^{(i)}}$ & the weight matrix between $\bm{F^{(i)}}$ and $\bm{F^{(i+1)}}$.\\
      $\bm{R_t}$ & the GCN-encoded representation for time $t$. \\ \hline
      $\bm{H_t}$ & the hidden states in the RNN for time $t$. \\
      $\bm{S}$ & the list of hidden states in the RNN over time.\\
      $\tau_t$ & the importance weight for $\bm{H_t}$.\\
      $\mathcal{F}_a(\cdot)$ & the fully-connected layer for deriving $\tau_t$.\\
      $\alpha_t$ & the normalized importance weight for $\bm{H_t}$.\\ 
      $\bm{c_u}$ & the final representation of the influencer $u$.\\ \hline
      $\bm{\hat{y}_u}$ & the predicted engagement score for the influencer $u$. \\
      $\mathcal{F}(\cdot)$ & the fully-connected layers for inferring $\bm{\hat{y}_u}$.\\ 
      \hline
    \end{tabular}}
    \caption{Summary of notations and their descriptions.}
    \label{tab:notations}
\end{table}

\subsection{Heterogeneous Information Networks}

To represent the dynamics of the engagement rates on a sequence of time, we build $k$ heterogeneous networks $\mathbb{G}= \lbrace G_1, G_2, \cdots, G_k\rbrace$ based on the influencers and other relevant entities.
Hence, $G_t$ can further characterize the relationships of influencers and their posting behaviors at time $t$.


\subsubsection{Heterogeneous Nodes and Embedded Features}
We build a heterogeneous network $G_t$ for time $t$ with four different types of nodes, including influencers, hashtags, image objects, and other users in social media.
Given an influencer $u$, we extract all of the hashtags $\lbrace h_i \rbrace_{i=1}^{a} \in H$ and mentioned users (i.e., user tags) $\lbrace v_j \rbrace_{j=1}^{b} \in V$ from posts $P^t_u$, where $a$ and $b$ indicate the number of extracted hashtags and mentioned users, respectively.
Note that the mentioned users can be either influencers, brands, or other normal users.
In addition, the categories of objects shown in the posted images $\lbrace o_k \rbrace_{k=1}^{c} \in O$ are also considered as nodes. 
Since each type of node has unique features, we denote the node features of influencers, mentioned users, hashtags, and object categories in images as $\bm{X_t^U}$, $\bm{X_t^V}$, $\bm{X_t^H}$, and $\bm{X_t^O}$, respectively.
We then represent embedded features of each node as $\bm{X_t} = [\bm{X_t^U}; \bm{X_t^V}; \bm{X_t^H}; \bm{X_t^O}]\in \mathbb{R}^{N\times d},$ where $N$ is the total number of all four types of nodes and $d$ is the number of embedded node features. 


\subsubsection{Edge Construction and Adjacency Matrix}
The edges in the heterogeneous network indicate the interactions between entities behind nodes.
For example, if an influencer mentioned the hashtag \texttt{\#makeup} and posted an image of cosmetic products, the influencer node will be connected to the node of the \texttt{\#makeup} hashtag and the node of the cosmetic image object.
Given a timestamp $t$, we make a sparse adjacency matrix $\bm{A_t}\in \mathbb{R}^{N\times N}$, where $A_{ij}^{t} = 1$ indicates a connection between the $i$-th and $j$-th nodes.

Finally, a set of $k$ heterogeneous networks $\mathbb{G}$ with the sets of node features and adjacency matrices can be constructed as follows: 
$$\mathbb{G}= \lbrace G_1, G_2, \cdots, G_k\rbrace,$$
where $G_t = \bm{(X_{t},A_{t})}$ indicates both the node embedded features $\bm{X_{t}}$ and the heterogeneous network structure $A_{t}$ at time $t$.

\subsection{Graph Convolutional Networks}

For the heterogeneous network $G_t$ of each time $t$, our proposed InfluencerRank applies Graph Convolutional Networks~(GCNs)~\cite{kipf2016semi} to generate node representations over time.
GCNs first generate a normalized adjacency matrix~$\bm{\hat{A_t}}$ by transforming the adjacency matrix~$\bm{A_t}$ with the diagonal degree matrix~$D$ as $\bm{\hat{A_t}} = \bm{D}^{-\frac{1}{2}}A_{t}\bm{D}^{-\frac{1}{2}}.$
GCNs then stack multiple GCN layers where each layer takes outputs of the previous layer and performs nonlinear transformation to propagate information through different layers.
The $i$-th layer in GCNs then outputs $\bm{F^{(i)}}\in \mathbb{R}^{N\times r}$ as follows:
$$\bm{F^{(i)}} = \sigma\left(\bm{\hat{A_t}} \bm{F^{(i-1)}} \bm{W^{(i-1)}} \right),$$
where $r$ is the number of hidden dimensions in GCNs, $\bm{F^{(i-1)}}$ is the outputs of the previous layer, $\bm{W^{(i-1)}}$ is a matrix of trainable weights, and $\sigma(\cdot)$ is a nonlinear activation function.
We use $\bm{X_t}$ for $\bm{F^{(0)}}$ as the input of the first GCN layer.
The final output of the GCNs $\bm{R_t}$ at time $t$ can be represented as follows:
$$\bm{R_t} = \left[\bm{F^{(1)}}, \bm{F^{(2)}}, \dots, \bm{F^{(e)}}\right],$$
where $e$ is the number of layers in GCNs.

Finally, we can obtain a sequence of GCN-encoded node representations, $[\bm{R_1}, \dots, \bm{R_k}]$, to implicitly represent the knowledge about influencers over time.

\subsection{Attentive Recurrent Neural Networks}

\subsubsection{Learning Graph Dynamics}
Based on the sequence of GCN-encoded node representations, $[\bm{R_1}, \dots, \bm{R_k}]$,  InfluencerRank applies Recurrent Neural Networks (RNNs) to the model framework. 
More specifically, we employ Gated Recurrent Units~(GRUs)~\cite{cho2014learning}, which use update gate and reset gate inside the unit to carry information flow over many time periods, to capture long-term temporal dependencies from the heterogeneous networks.
Each GRU takes hidden states from the previous unit and the GCN representations as input and then outputs hidden states of the current time.
More formally, the hidden states at time $t$, $H_{t}$ is computed as follows:
\begin{equation} \label{eq_2}
H_{t} = (1-z_{t})H_{t-1} + z_{t}\tilde{H_{t}}
\end{equation}
where $z_{t}$ is an update gate at time $t$ and $\tilde{H_{t}}$ is the candidate state at time $t$. The candidate state is updated as follows:
\begin{equation} \label{eq_3}
\tilde{H_{t}} = tanh(W \cdot [r_{t} \odot H_{t-1},R_{t}])
\end{equation}
where $r_{t}$ is an reset gate at time $t$, $\odot$ is an element-wise multiplication, and $R_{t}$ is the GCN representations at time $t$.
Finally, InfluencerRank obtains the whole states of GRUs as follows:
$$\bm{S} = \left[\bm{H_{1}}, \bm{H_{2}}, \dots, \bm{H_{k}}\right].$$

\subsubsection{Attention over Time}
To acquire the final influencer representations, InfluencerRank applies the attention mechanism~\cite{bahdanau2014neural} to the whole state embeddings derived from GRUs~$\bm{S}$.
The attention mechanism allows InfluencerRank to learn the dynamics of the engagement rates by taking into account the importance of time periods.

For each timestep $t$, InfluencerRank estimates the importance weight of the corresponding state embedding by applying a projection as:
\begin{equation} \label{eq_4}
\tau_t = \tanh\left(\mathcal{F}_{a}\left(\bm{H_t}\right)\right)
\end{equation}
where $\mathcal{F}_{a}(\cdot)$ is a fully-connected layer; $tanh(\cdot)$ is the activation function.
We then compute the weights of each timestep by using a softmax function as:
\begin{equation} \label{eq_5}
\alpha_t = \frac{\exp(\tau_t)}{ \sum_{i=1}^{k} \exp(\tau_i)}
\end{equation}
Finally, InfluencerRank derives the ultimate representation of candidate influencers by using the weighted sum as follows:
\begin{equation} \label{eq_6}
\bm{c} = \sum_{i=1}^{k} \alpha_i \cdot \bm{H_{i}}
\end{equation}

\subsection{Engagement Score Estimation}

For an influencer $u$, InfluencerRank takes the corresponding ultimate representation $\bm{c_u}$ as the input and then predicts an engagement score  $\hat{y_u}$ that is proportional to the engagement rate $E^k_u$ as follows:
\begin{equation} \label{eq_7}
\bm{\hat{y_u}} = \mathcal{F}_c\left(\text{ReLU}\left(\mathcal{F}_b\left(\bm{c_u}\right)\right)\right)
\end{equation}
where a non-linear transformation is carried out in a fully-connected layer $\mathcal{F}_{b}(\cdot)$ with the ReLU activation function and the engagement rate is estimated in another fully-connected layer $\mathcal{F}_{c}(\cdot)$.

\subsubsection{List-wise Ranking and Optimization}
InfluencerRank treats the task as a ranking problem and optimizes the ranking performance with a list-wise learning-to-rank framework~\cite{xia2008listwise}.
Suppose $Z$ is the set of features for influencers to be ranked; $Y$ is the space of all possible rankings.
During training, we sample $m$ labeled influencers from the whole training space as an i.i.d. candidate ranked list $S = \left\lbrace (Z_i, \bm{y_i})\right\rbrace^m_{i=1} \sim P_{ZY},$ where $P_{ZY}$ is the unknown target joint probability distribution of $Z$ and $Y$.
Therefore, the corresponding loss $\mathcal{L}_S$ can be considered as:
\begin{equation} \label{eq_8}
\mathcal{L}_S(\bm{\hat{y}}) = \frac{1}{m} \sum_{i=1}^m \bm{l}(\bm{\hat{y}}(\bm{Z_i}), \bm{y_i})
\end{equation}
where $\bm{l}(\bm{\hat{y}}(\bm{Z_i}), \bm{y})$ is the 0-1 loss between $\bm{\hat{y}}(\bm{Z_i})$ and the rank in $\bm{y}$; $\bm{y_i}$ denotes the ground-truth ranking such that
\begin{equation} \label{eq_9}
L(\bm{\hat{y}}(\bm{Z_i}), \bm{y}) = \left\lbrace \begin{array}{ccl}
     1 & , \text{ if } \bm{\hat{y}}(\bm{Z_i}) \neq \bm{y}  \\
     0 & , \text{ if } \bm{\hat{y}}(\bm{Z_i}) = \bm{y}  \\
\end{array}\right.
\end{equation}

\subsection{Node Features}
\label{sec:features}

In this subsection, we describe node features in the heterogeneous network.
To understand the relationship between the engagement rate of an influencer and the characteristics of the corresponding influencer, we introduce six types of node features, including node type, profile, image, text, posting, and reaction features.
Note that most of the features are only applicable for influencer nodes while the remaining nodes (e.g., hashtags, image objects) hold zeros for the inapplicable features.
For the feature engineering, we deploy the average, median, minimum, and maximum values for the features that need to be aggregated with statistics.

Here, we briefly introduce the six categories of features used in this paper as follows.
\begin{itemize}
    \item \textbf{Node type features.}
    The one-hot coded feature that indicates one of the four node types, including influencers, other users, hashtags, and image objects.

    \item \textbf{Profile features.}
    For each influencer node, we exploit the numbers of followers, followees, and posts which are the most commonly used metrics to measure user influence in social networks~\cite{riquelme2016measuring}.
    Additionally, we consider a category of influencers from eight influencer categories defined in the previous study~\cite{kim2020profile}.

    \item \textbf{Image features.}
    The previous study~\cite{gelli2015image} showed that the characteristics of images on social media posts affect its popularity.
    In addition to the image objects which are considered as nodes in the heterogeneous network, we add the attributes of visual perception of the images to understand how influencers create images.
    We compute the brightness, colorfulness~\cite{hasler2003measuring}, and color temperature of the posted images based on their RGB values.

    \item \textbf{Text features.}
    To understand how textual usage of influencers affects the engagement rate, we retrieve various text features. More specifically, we use the numbers of hashtags, user tags, and emojis that are widely used functions on social media, and the length of captions which can represent how much detailed information is in the caption~\cite{hessel2017cats}.
    Moreover, we also calculate the sentiment scores of captions to learn how positive or negative emotions are carried through the captions by using VADER~\cite{gilbert2014vader}.

    \item \textbf{Posting features.}
    The features in this category can provide information about how influencers use social media from various aspects. 
    We first exploit the portion of the number of posts in one of the ten post categories~\cite{kim2020profile} to the total number of posts to understand the posting behavior of influencers.
    In addition to the post category rate feature, we also examine the portion of the number of advertising posts to the total posts published by an influencer; posting too many paid advertisements can show negative impacts on the popularity~\cite{evans2017disclosing,yang2019influencers}. 
    We also consider the feedback rate and posting interval, which are the measures of the interaction with their followers and activeness, respectively. The feedback rate is calculated as the ratio of the number of posts that contain the influencers' responses to the user comments to the number of total posts.
    The posting interval is the average time gap between posts that are in chronological order.

    \item \textbf{Reaction features.}
    We use the user comments to generate the user reaction feature. Specifically, we compute the sentiment scores of comments that are written by audiences of the influencers' posts. 
    Note that we do not consider the number of likes and comments as node features since it can directly imply the engagement rates of influencers.

\end{itemize}

\section{Experiments}\label{sec:experiment}

In this section, we conduct experiments to evaluate the performance of InfluencerRank compared with other baseline methods. We also analyze the experimental results to understand the importance of each feature to find influencers with high engagement rates.


\subsection{Experimental Dataset}

\subsubsection{Dataset Construction} 
To evaluate the proposed \textit{InfluencerRank}, we use the Instagram influencer dataset~\cite{kim2020profile}. 
The dataset includes profiles of influencers, and their posts, including both images and all meta-data.
We only keep the posts that were published in the range of January 1st, 2017 and December 31st, 2017, to build temporal influencer networks. 
As a result, the dataset consists of 18,397 influencers and 2,952,075 posts.
For the experiments, we split the dataset into the training dataset, which contains posts from January to November, and the testing dataset that contains posts published in December.

\subsubsection{Heterogeneous Network Construction}
To build the temporal heterogeneous networks, we first divide the whole dataset into 12 subsets by one-month intervals. Note that we conduct experiments to analyze ranking performances across different temporal window sizes, thereby having the proper time intervals.
We then extract all hashtags and user tags from the post captions and detect objects from the images.
As a consequence, 1,151,082 unique hashtag nodes, 532,468 other user nodes, and 1,000 image object nodes are found across the networks and connected to the corresponding influencer nodes.
To further reduce noises in the dataset, we remove every auxiliary node (i.e., hashtags, other users, and image objects) with only a single edge while edges with normalized frequencies less than 0.01 are also discarded.
After the pruning process, 18,397 influencers, 20,744 other users, 67,695 hashtags, and 996 image objects are in the networks (i.e., 107,832 nodes), and a total of 15,090,225 edges remain across the networks.


\subsection{Experimental Settings}

\subsubsection{Evaluation Metrics}
Based on the definition, we first compute the engagement rates for all influencers in across all timesteps as the ground truths.
Note that the average engagement rate is 0.038 and the median engagement rate is 0.029.
We utilize two metrics to evaluate the performance of ranking influencers.
\begin{itemize}[leftmargin=*]
    \item \textit{Normalized Discounted Cumulative Gain} (NDCG) \cite{jarvelin2017ir}: First, we divide all of the influencers into six groups with different thresholds on the engagement rates and relevance levels from 0 to 5. Table~\ref{tab:label_stat} further shows the statistics of influencers in the dataset across different relevance levels and criteria for the engagement rates. We then treat the relevance levels as ground truths to evaluate the ranking performance with the metric of NDCG.
    \item \textit{Rank-Biased Precision} (RBP) \cite{moffat2008rank}: To avoid losing valuable information while converting the engagement rates to the six relevance levels, we directly use the engagement rates with the metric of RBP. We set the probability $p$ as 0.95 to measure rank quality.
\end{itemize}

\begin{table}[!t]
    \renewcommand{\arraystretch}{1.0}%
    \small
    \centering
    \begin{tabular}{|c|c|c|}\hline
         Relevance & Engagement rate $E(\cdot)$ & Number of Influencers \\ \hline
         5 & $E(\cdot) \geq 0.10$ & 1,274~(6.92\%)\\
         4 & $0.10 > E(\cdot) \geq 0.07$ & 1,678~(9.12\%)\\
         3 & $0.07 > E(\cdot) \geq 0.05$ & 2,321~(12.62\%)\\
         2 & $0.05 > E(\cdot) \geq 0.03$ & 4,509~(24.51\%)\\
         1 & $0.03 > E(\cdot) \geq 0.01$ & 6,882~(37.41\%)\\
         0 & $0.01 > E(\cdot)$ & 1,734~(9.42\%) \\ \hline
    \end{tabular}
    \caption{Statistics of influencers in the dataset across different relevance levels and criteria for the engagement rates.}
    \label{tab:label_stat}
\end{table}

\subsubsection{Implementation Details}
For the hyperparameter tuning, we use a validation set which contains posts published by the 18,397 influencers in January 2018. 
Since our model is optimized with the validation set, we can avoid potential information leakage from the testing set.
In order to fine-tune the model, we train the proposed neural network with different sets of parameters, including the number of GCN layers and its dimensions, the size of influencer list for ranking, batch size, learning rate, and dropout probability. 
After tuning the model, we set the numbers of dimensions of the graph embeddings and GCN features as 128, and the number of GCN layers as 2.
Each batch contains 1,024 lists of influencers, and each list includes 10 randomly selected influencers for list-wise learning.
The learning rate and the dropout probability are set as 0.001 and 0.5, respectively. 
 
\begin{table*}[t]
\renewcommand{\arraystretch}{1.0}%
\centering
\resizebox{0.55\linewidth}{!}{
\begin{tabular}{|c||c|c|c|c|c|c|c|} 
 \hline
 \multirow{2}{*}{Method} & \multirow{2}{*}{\textit{RBP}} & \multicolumn{5}{c|}{\textit{NDCG@K}} & Time\\ \cline{3-7}
  & & 1 & 10 & 50 & 100 & 200 & (sec) \\ \hline \hline
 UP & 0.025 & 0.800 & 0.436 & 0.413 & 0.406 & 0.368 & 347\\
 PP & 0.028 & 1.000 & 0.519 & 0.465 & 0.442 & 0.425 & \textbf{295}\\
 UA & 0.024 & 0.800 & 0.518 & 0.494 & 0.438 & 0.436 & 330\\
 \hline
 LN & 0.026 & 1.000 & 0.610 & 0.511 & 0.465 & 0.441 & 481\\
 LM & 0.031 & 1.000 & 0.648 & 0.546 & 0.493 & 0.477 & 563\\ 
 \hline
 \textit{GCRN} & 0.028 & 1.000 & 0.629 & 0.557 & 0.513 & 0.467 & 612\\ 
 \textit{DeepInf} & 0.031 & 1.000 & 0.697 & 0.567 & 0.549 & 0.512 & 525\\ 
 \textit{CasCN} & 0.033 & 1.000 & 0.751 & 0.645 & 0.572 & 0.543 & 1109\\
 \textit{EGCN} & 0.038 & 1.000 & 0.812 & 0.679 & 0.616 & 0.577 & 1483\\
 
 \hline
 \textit{InfluencerRank} & \textbf{0.043} & \textbf{1.000} & \textbf{0.864} & \textbf{0.720} & \textbf{0.661} & \textbf{0.614}  & 648\\ \hline
\end{tabular}}
\caption{\textit{RBP}, \textit{NDCG@K} scores, and training time of the proposed InfluencerRank and the nine baseline methods.}
\label{table:result}
\end{table*}


\subsubsection{Baseline Methods}

We compare the performance of \textit{InfluencerRank} with nine baseline methods in three different categories, including \textit{User}, \textit{Ranking}, and \textit{Graph}.

\begin{itemize}
    \item \textit{\textbf{User baselines}}: The baseline methods in this category exploit information on social media to measure the popularity of users with certain features. Since user popularity is often to be considered as an important factor in influencer hiring~\cite{de2017marketing,lou2019influencer,casalo2018influencers}, we develop three baseline methods including user popularity~(UP)~\cite{bakshy2011everyone}, post popularity~(PP)~\cite{mazloom2016multimodal}, user activity~(UA)~\cite{li2011discovering}.
    \item \textit{\textbf{Ranking baselines}}: We consider two ranking models, \textit{ListNet} (LN)~\cite{cao2007learning} and \textit{LambdaMART}~(LM)~\cite{burges2007learning}. The ranking baseline methods use the same features as our proposed model.
    \item \textit{\textbf{Model baselines}}: The baseline methods in this category implement graph neural networks~(GNNs) based learning models. Note that the applied features of \textit{InfluencerRank} and graph baselines are identical so that we can fairly evaluate the model novelty and capability for the ranking task. We develop GCRN~\cite{seo2018structured}, DeepInf~\cite{qiu2018deepinf}, CasCN~\cite{chen2019information}, and EGCN~\cite{pareja2020evolvegcn}.
\end{itemize}

\subsection{Experimental Results}

Table~\ref{table:result} shows RBP, NDCG scores, and training time of InfluencerRank and the nine baseline methods for discovering influencers with high engagement rates.
All of the three methods in the user baselines, which exploit the social media features, obtain low ranking results.
This suggests that only considering social media features is insufficient to discover effective influencers.
The ranking baseline methods, on the other hand, show better ranking performance compared to the user baseline methods since they use our proposed features.
It demonstrates that our proposed features are very useful to capture the characteristics of influencers, thereby discovering effective influencers.
Next, most of the graph baseline methods outperform the user baselines and ranking baselines.
More specifically, among the graph baseline methods, GCRN~\cite{seo2018structured} shows limited ranking performance improvement since it only resorts to temporal-spatial structures of graphs without taking into account the node features.
DeepInf~\cite{qiu2018deepinf} demonstrates better ranking performance than GCRN by exploiting the graph convolutional networks that take advantage of the network structures of different entities while the features in different aspects provide sufficient knowledge to describe both influencers and other entities in the graph.
Both CasCN~\cite{chen2019information} and EGCN~\cite{pareja2020evolvegcn} further improve performance by applying recurrent neural networks to adjacency matrices of temporal graphs.
This suggests that the learning dynamics of graph structures with node features over time is beneficial in discovering effective influencers.

Finally, our proposed approach, InfluencerRank, outperforms all of the baseline methods.  
This is because our model derives informative influencer representations over time by using the graph convolutional networks and the attentive neural network, and effectively learns the dynamics of influencer characteristics and engagement rates.
The results also show that InfluencerRank is able to learn the latent influencer representations in a reasonable amount of training time compared to the other baseline methods.
CasCN and EGCN, on the other hand, have significantly longer training time than InfluencerRank.
This is probably because the proposed framework successfully learns the importance of hidden states in the RNNs by applying attention while other baseline methods combine RNNs with GCNs without taking the importance of each temporal graph into account.


\section{Analysis and Discussions}
In this section, we conduct six analyses to understand the importance of (i) the temporal window size, (ii) the temporal information, (iii) the model components, (iv) the type of RNNs, (v) the heterogeneous networks, and (vi) the input features. 
We then evaluate the performance of InfluencerRank on various sets of influencers which are grouped by the size of audiences.

\subsection{Analysis on Temporal Window Size}\label{sec:temporal_window_size}
\begin{figure}[h]
\centering
\includegraphics[width=0.9\linewidth]{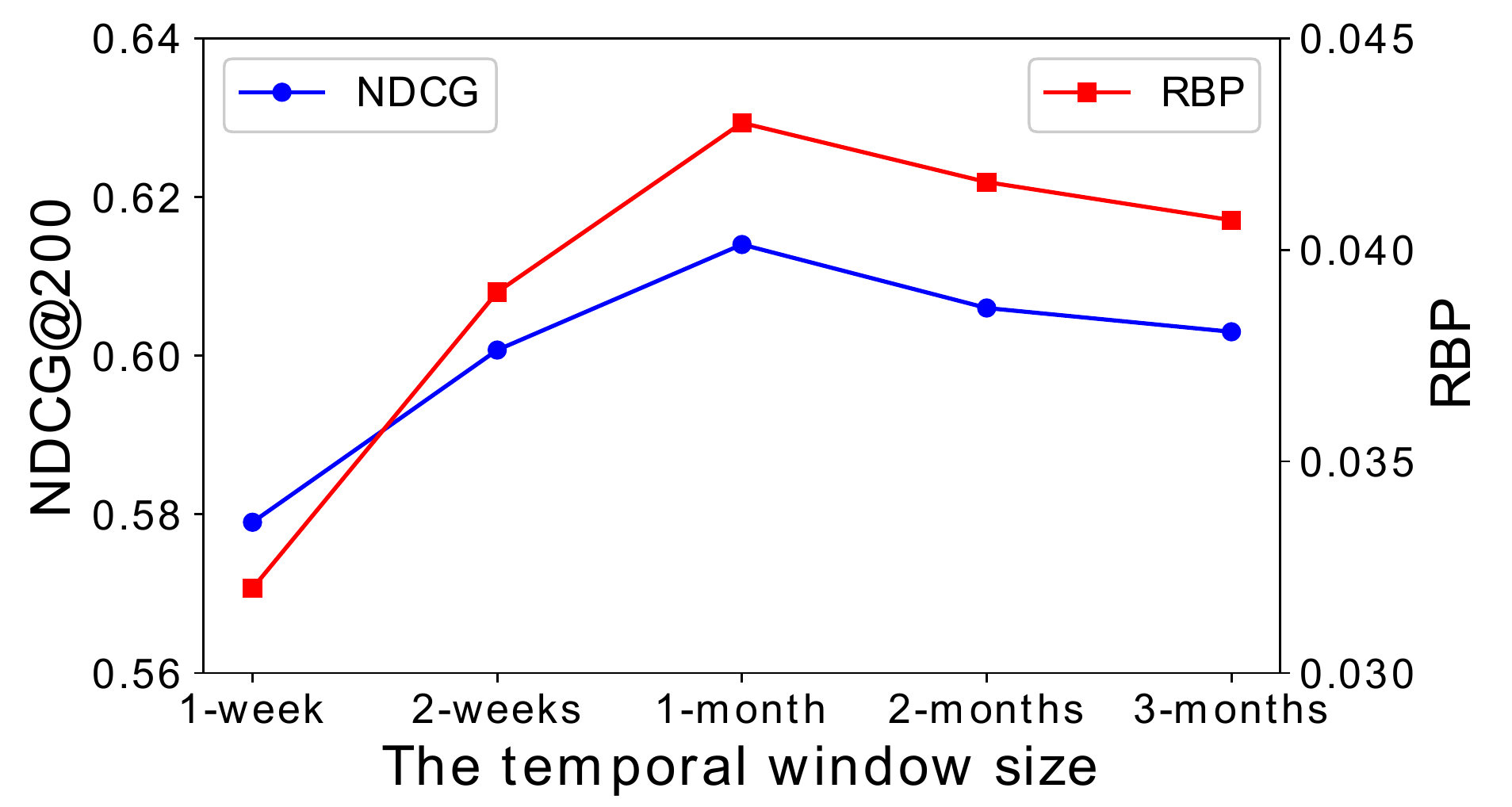}
\caption{Ranking performance over different temporal window sizes. InfluencerRank trained with the 1-month window size shows the best performance.}
\label{fig:temporal_window_size}
\end{figure}

We first investigate the effect of different temporal window sizes for heterogeneous network construction.
To that end, we split the training dataset which has posts in 11 months period into sub-datasets by five different temporal window sizes including 1 week, 2 weeks, 1 month, 2 months, and 3 months.
Note that we use the same testing dataset across the five different window sizes for consistent performance comparison.
The RBP and NDCG@200 scores of the InfluencerRank over the different temporal window sizes are shown in Figure~\ref{fig:temporal_window_size}.
We find that the model trained with the networks divided by 1-month intervals shows the best ranking performance whereas the model trained with the 1-week temporal window has the lowest ranking scores.
InfluencerRank loses 5.7\% performance on NDCG when the model is trained with the 1-week window size compared to the model trained with the 1-month window size.
This suggests that the heterogeneous networks of the models, which are trained with temporal window size shorter than 1-month, have insufficient information to learn the dynamics of engagement rates.
We also observe that the ranking performance gradually decreases while we use the longer temporal window size.
This implies that the model trained with a large window size fails to take into account the variance of engagements by using the average like counts of all posts in each sub-dataset.
The analysis results also demonstrate that learning the temporal dynamics of the engagement rates is very important to find effective influencers.

\subsection{Analysis on Temporal Information}
\begin{figure}[h]
\centering
\includegraphics[width=0.9\linewidth]{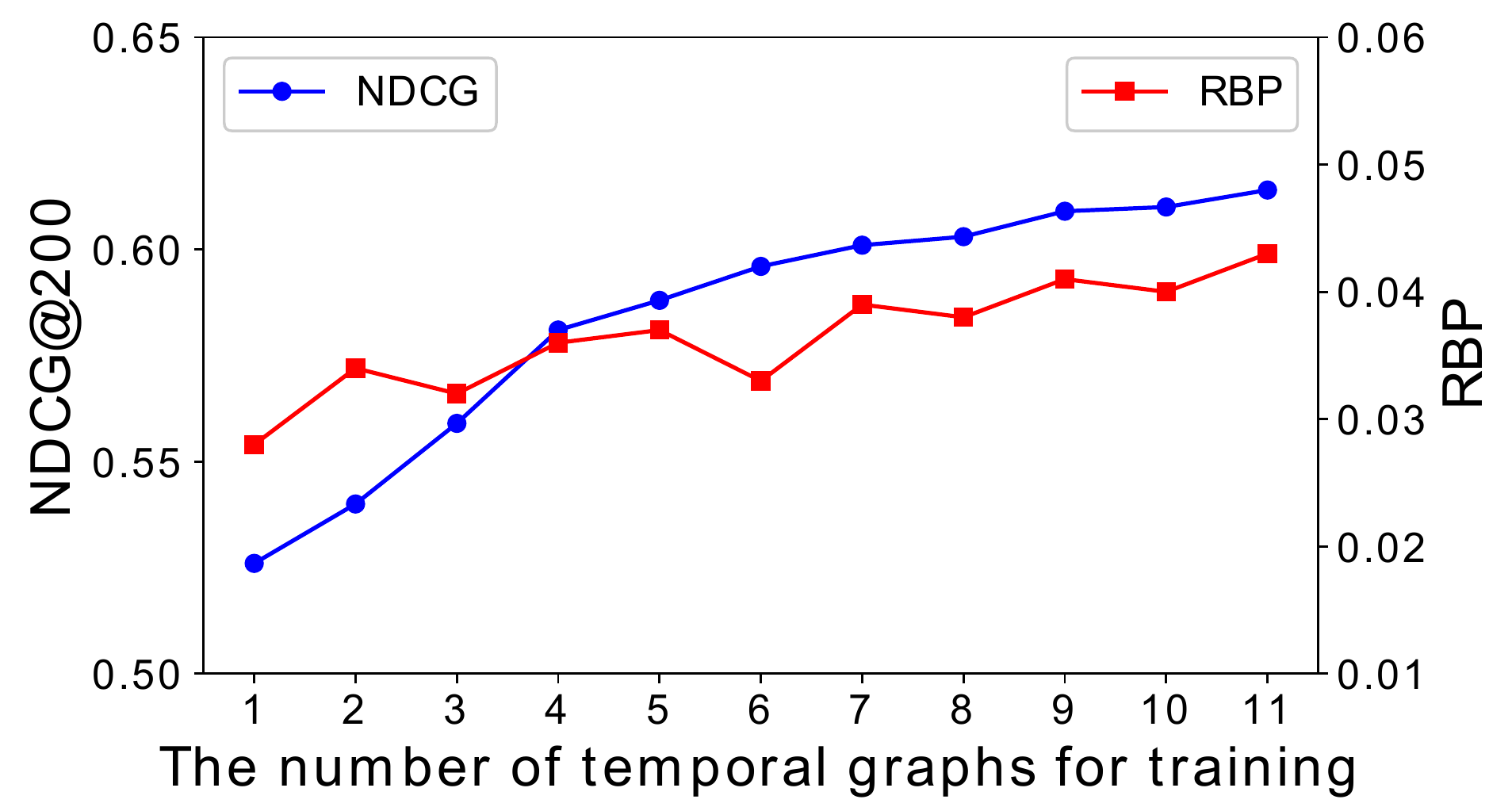}
\caption{Ranking performance on different lengths of timestamps. InfluencerRank achieves higher ranking scores with longer history.}
\label{fig:temporal_information}
\end{figure}

We next evaluate the ranking performance of the proposed model by using the different number of temporal input networks for training the model.
Figure~\ref{fig:temporal_information} shows RBP and NDCG@200 scores of the InfluencerRank over the number of temporal graphs.
Note that the model uses the most recent temporal graphs.
For example, a model trained with two temporal graphs learns two networks in October and November in our dataset.
We observe that the performance significantly drops when InfluencerRank obtains insufficient historical information.
InfluencerRank loses 15\% performance on NDCG if the model uses only one graph compared to the model that considers all temporal graphs.
The result confirms that only considering the most recent network degrades the performance since the engagement rates of influencers vary over time.
We also find that as the number of temporal networks increases, the model has gradually less performance gain.

\subsection{Analysis on Model Components}
\begin{figure}[h]
\centering
\includegraphics[width=.9\linewidth]{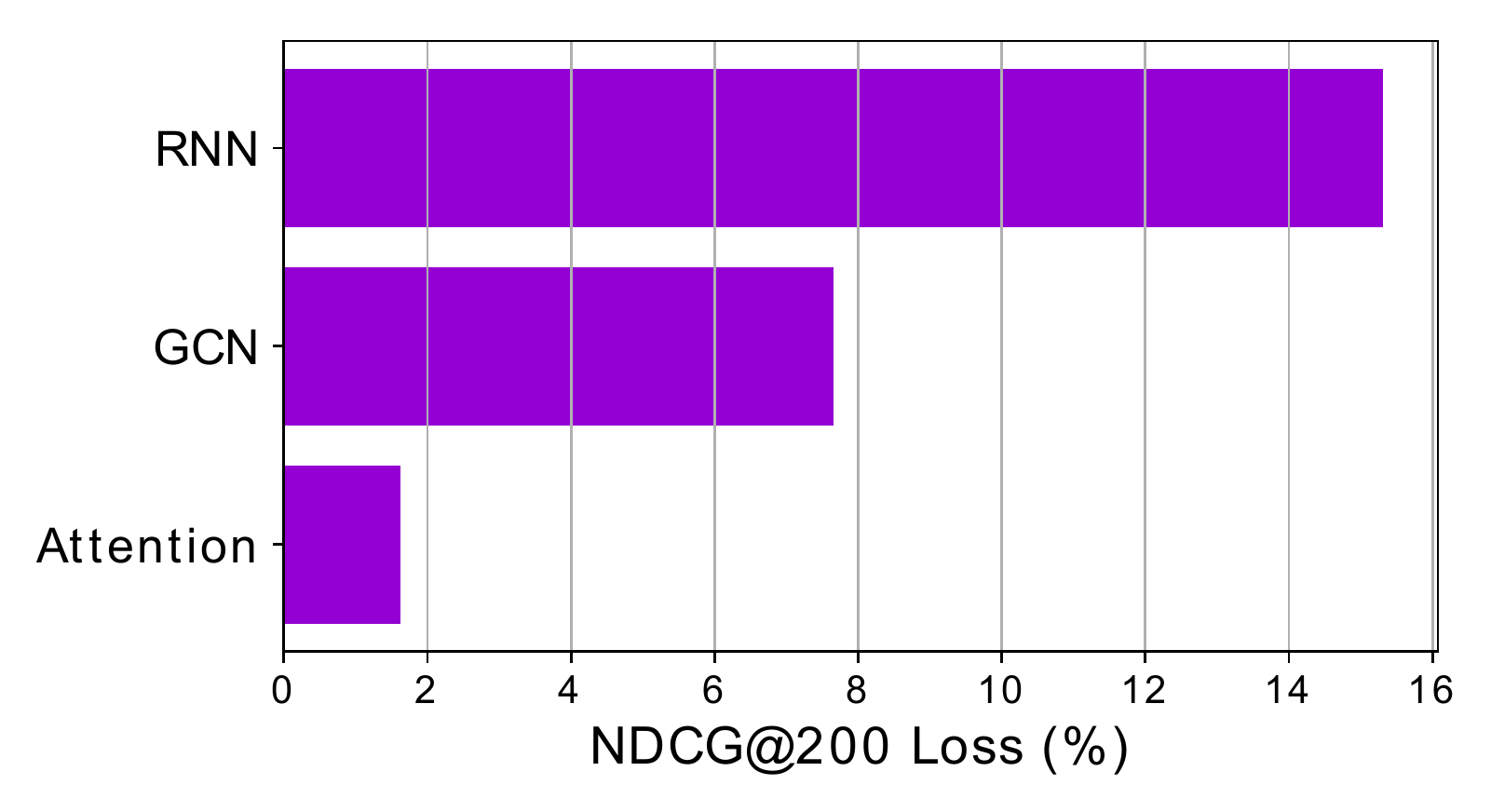}
\caption{Performance losses after removing each of the model components. The RNNs are the most important component to discover effective influencers.}
\label{fig:model_components}
\end{figure}

The proposed model consists of three major components, including the graph convolutional networks, the recurrent neural networks, and the attention network. 
We conduct an ablation study by excluding each component from the model framework to understand the importance of model components.
Figure~\ref{fig:model_components} shows the performance losses of NDCG@200 scores over the three model components.
We find that the model which excludes the RNN component has significant performance loss compared to the full model.
This suggests that disregarding to learn sequential temporal information leads to performance degradation since engagement rates of an influencer change over time.
The model that discards the GCN component also shows large performance loss.
This is because the model fails to learn structural information with embedded node features.
This demonstrates that learning social relationships of influencers with other users, tags, and image objects plays an important role in discovering effective influencers.
We observe that the attention component has relatively less impact on the performance than other model components whilst it still enhances the model by considering the importance of temporal graph embeddings.

\subsection{Analysis on Recurrent Neural Networks}\label{sec:rnn_analysis}

\begin{figure}[h]
\centering
\includegraphics[width=.9\linewidth]{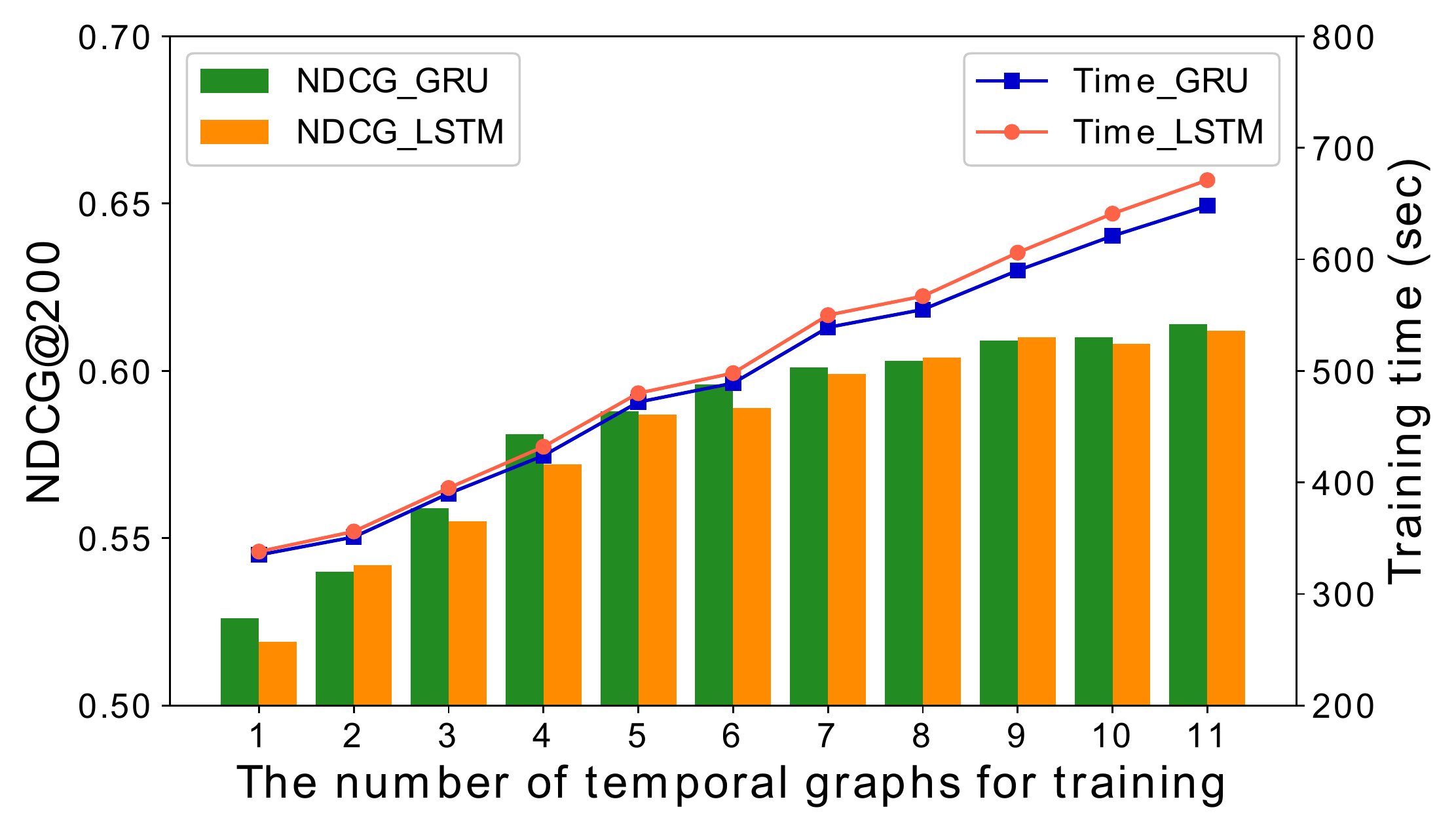}
\caption{NDCG scores and training times of InfluencerRank trained with GRU and LSTM on different number of temporal graphs. InfluencerRank with GRU tends to have better ranking performance and shorter training time than the model with LSTM.}
\label{fig:gru_lstm}
\end{figure}

In the proposed InfluencerRank framework, we employ gated recurrent units~(GRUs)~\cite{cho2014learning} for the recurrent neural networks.
However, the GRU can be replaced with a long short-term memory~(LSTM)~\cite{hochreiter1997long}. 
To make a design decision which recurrent architecture to employ, we train InfluencerRank with GRU and LSTM.
Figure~\ref{fig:gru_lstm} shows NDCG@200 scores and training times of InflurncerRank with two RNN architectures on different number of temporal graphs.
We observe that no significant difference in the NDCG scores of models using GRU and LSTM, but the model with GRU tends to have slightly higher scores. 
The results also show that InfluencerRank with GRU has shorter training times than LSTM across the different number of temporal graphs.
More specifically, the time difference gradually increases as the number of temporal graphs for training increases.
Note that GRU is 1\% faster than LSTM when the model only takes one temporal graph and 3.4\% faster when the model uses 11 graphs for training.
GRU shows better performance than LSTM in our task and that is probably because GRU has simpler network than LSTM and also benefits from the short input sequence length.

\subsection{Analysis on Heterogeneous Networks}
We study the importance of the proposed heterogeneous network to find effective influencers.
To understand the importance of individual auxiliary node type, we train InfluencerRank with the network without the type of auxiliary node.
Table~\ref{table:network} shows the RBP and the NDGC scores of InfluencerRank with different types of networks.
The results show that the model trained with all types of nodes achieve higher RBP and NDCG scores than the other models that exclude a type of auxiliary nodes. 
This confirms that the graphical structure in InfluencerRank helps improve performance in finding effective influencers.
We also observe that NDCG scores of the model without the image object nodes are lower than that of the model excluding hashtags and other user nodes.
Note that excluding the image object nodes drops the performance of NDCG@200 by 4.1\%, whereas excluding hashtag and other user nodes only drops the score by 2.8\% and 1.9\%.
This is probably because each image object node can densely connect a large number of similar influencers together as it has a greater number of edges than a hashtag node and a user node.

\begin{table}[h]
\renewcommand{\arraystretch}{1.0}%
\centering
\small
\resizebox{\linewidth}{!}{
\begin{tabular}{|c||c|c|c|c|c|c|} 
 \hline
 \multirow{2}{*}{Node Type} & \multirow{2}{*}{\textit{RBP}} &\multicolumn{5}{c|}{\textit{NDCG@K}} \\ \cline{3-7}
  & & 1 & 10 & 50 & 100 & 200 \\ \hline \hline
 All nodes                          & 0.043 & 1.000 & 0.864 & 0.720 & 0.661 & 0.614 \\ \hline
 $\bm{U}$, $\bm{V}$, $\bm{H}$       & 0.039 & 1.000 & 0.848 & 0.704 & 0.648 & 0.589 \\ \hline 
 $\bm{U}$, $\bm{H}$, $\bm{O}$       & 0.040 & 1.000 & 0.863 & 0.716 & 0.654 & 0.602 \\ \hline 
 $\bm{U}$, $\bm{V}$, $\bm{O}$       & 0.037 & 1.000 & 0.861 & 0.708 & 0.657 & 0.597 \\ \hline 
\end{tabular}}
\caption{Rank evaluation on the different network structures. Note that $\bm{U}$, $\bm{V}$, $\bm{H}$, and $\bm{O}$ represent influencer, other mentioned user, hashtag, and image object nodes in the network, respectively.}
\label{table:network}
\end{table}




\subsection{Analysis on Node Features}
The benefit of using GCNs comes from considering network structure information with node features. To understand the importance of node features, we first evaluate the performance of the model that excludes all node categories. 
The model without the whole node features significantly drops the ranking quality; the loss of NDGC@200 of the model without node features is 21.99\%.

\begin{figure}[h]
\centering
\includegraphics[width=.9\linewidth]{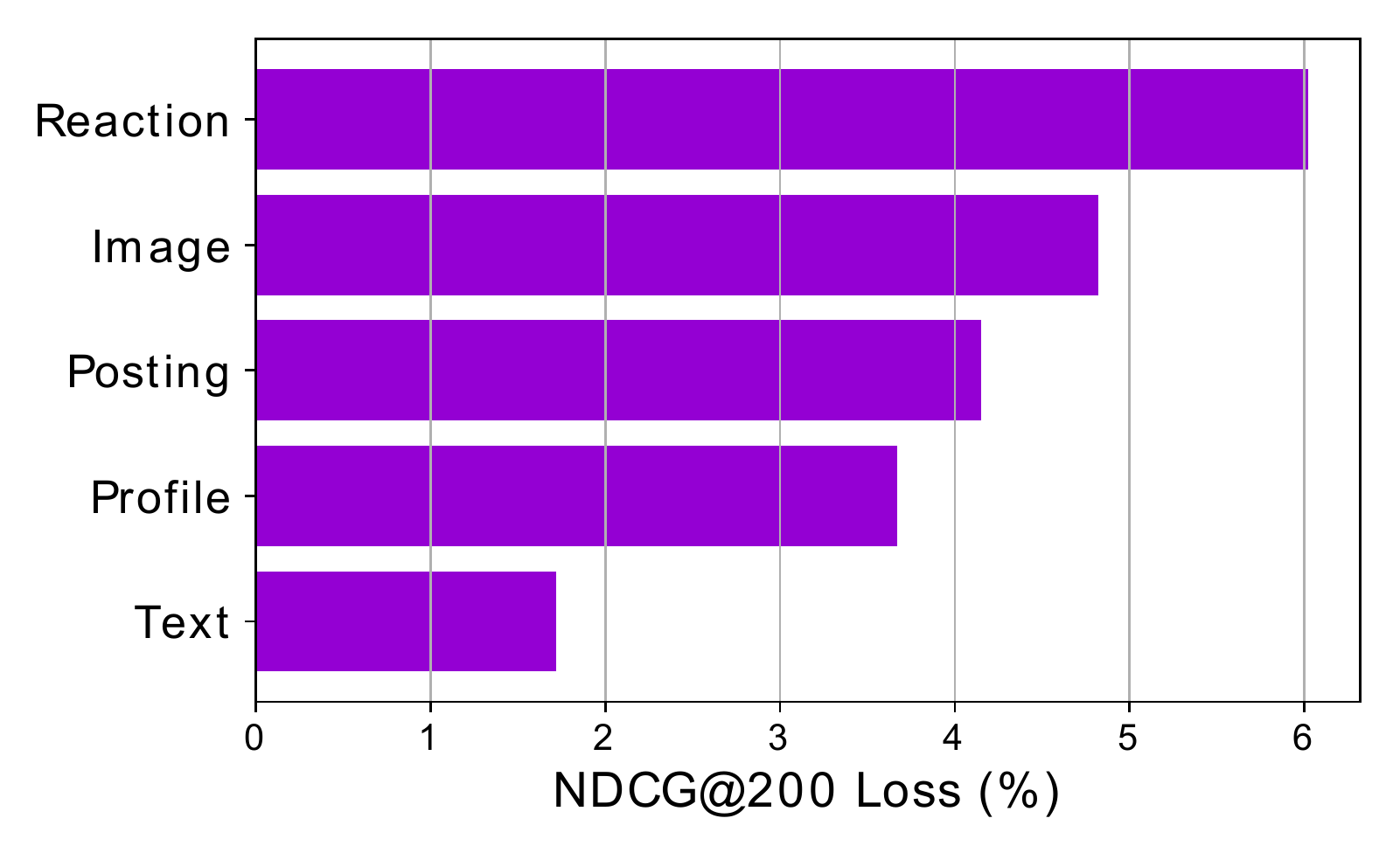}
\caption{Performance losses of NDCG@200 over node feature categories. The image features which represent visual perception and the reaction features which include sentiment scores of user comments have more impact on the effective influencer discovery than other types of features.}
\label{fig:node_features}
\end{figure}

We then investigate the performance of InfluencerRank with variant sets of node features to study the importance of each category of the node features.
Figure~\ref{fig:node_features} shows the performance loss of NDCG scores of the models trained with the node features excluding one particular node category against the full model as the leave-one-out analysis.
The results reveal that the reaction feature category, which contains the sentiment scores of user comments on the posts, is more important than other categories to identify effective influencers. 
This indicates that the audience may show distinct reactions to influencers with high engagement rates.
The image category, which includes the visual perception of images (e.g., brightness, colorfulness), also has higher loss values than other node feature categories.
This suggests that influencers with high engagement rates may have different visual characteristics from other influencers.
On the other hand, the text category, including the number of hashtags, user tags, emojis in a caption, and the sentiment scores of the caption, have the least impact to discover effective influencers. 
Although the statistical features to represent textual characteristics of influencers' posts have less impact than other features, InfluencerRank can improve the ranking performance by taking hashtags and user tags into account to the network structure.


\subsection{Influencer Follower Size}
In the influencer marketing industry, influencers are often divided into subgroups by the number of followers since it directly refers to the size of potential customers and hiring cost~\cite{de2017marketing}. For example, companies with a sufficient marketing budget can hire influencers who are followed by millions of people while small retailers may collaborate with influencers with a small number of followers.
Therefore, we evaluate the performance of InfluencerRank over groups of influencers with different sizes of followers.
Although there are no standard criteria to classify influencers based on the number of followers, we utilize the following thresholds which are the generally accepted numbers to divide the influencers into three groups\footnote{\url{http://www.mattr.co/pros-cons-micro-macro-mid-level-influencers/}}. 
Influencers who are followed by less than 20,000 followers are classified as the \textit{Micro influencers}. The \textit{Mid-level influencers} have followers between 20,000 and 100,000, and \textit{Macro influencers} have more than 100,000 followers.
In our dataset, around 30\% of influencers are the micro-influencers, 45\% of them are the mid-level influencers, and the remaining 25\% influencers are the macro-influencers. 
To evaluate the performance under the same conditions, we randomly select multiple sets of 1,000 influencers from each category and run the experiment 10 times.

\begin{figure}[!t]
\centering
\subfigure[NDCG@50]{\includegraphics[width=.69\linewidth]{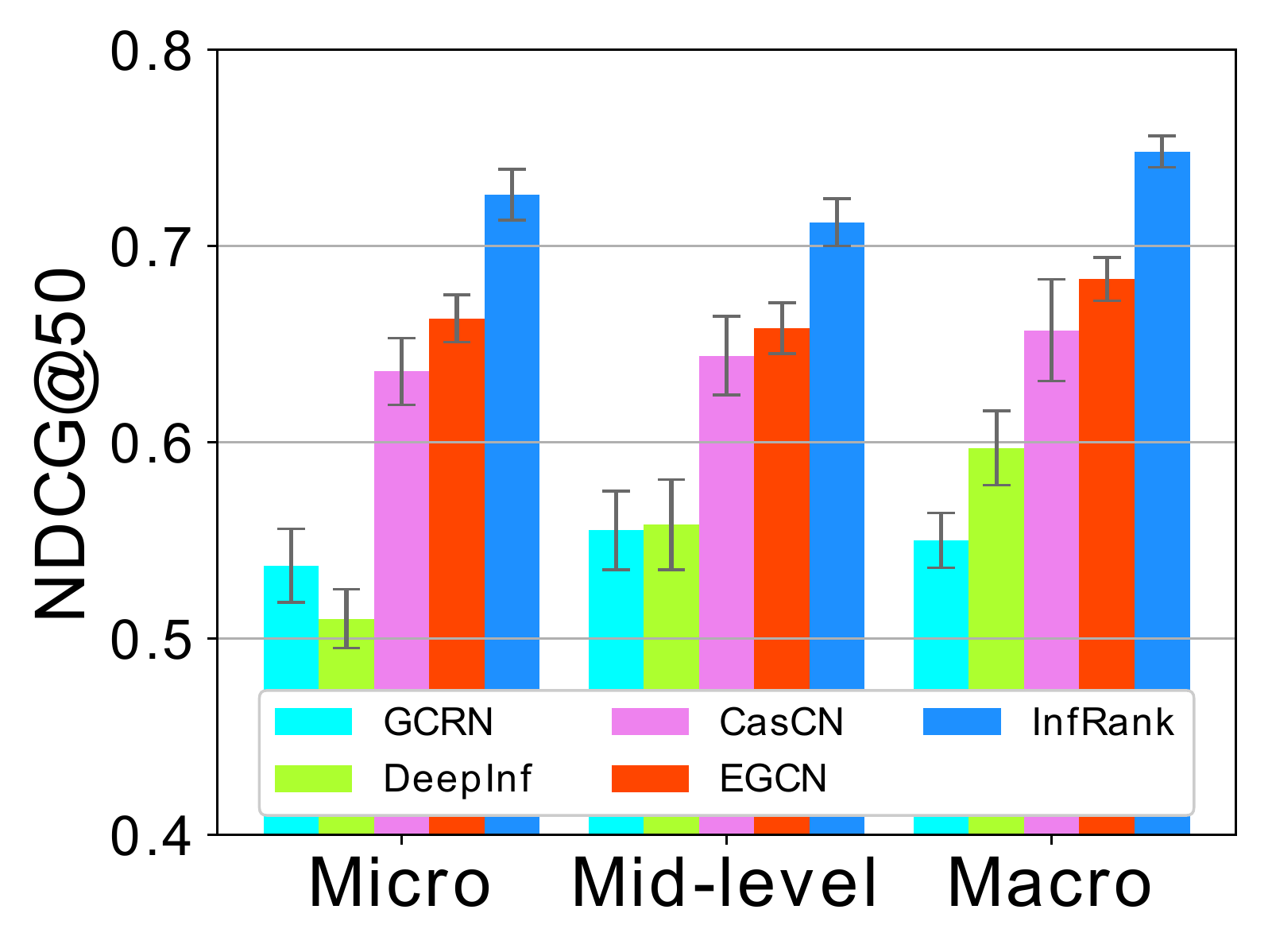}}
\subfigure[NDCG@200]{\includegraphics[width=.69\linewidth]{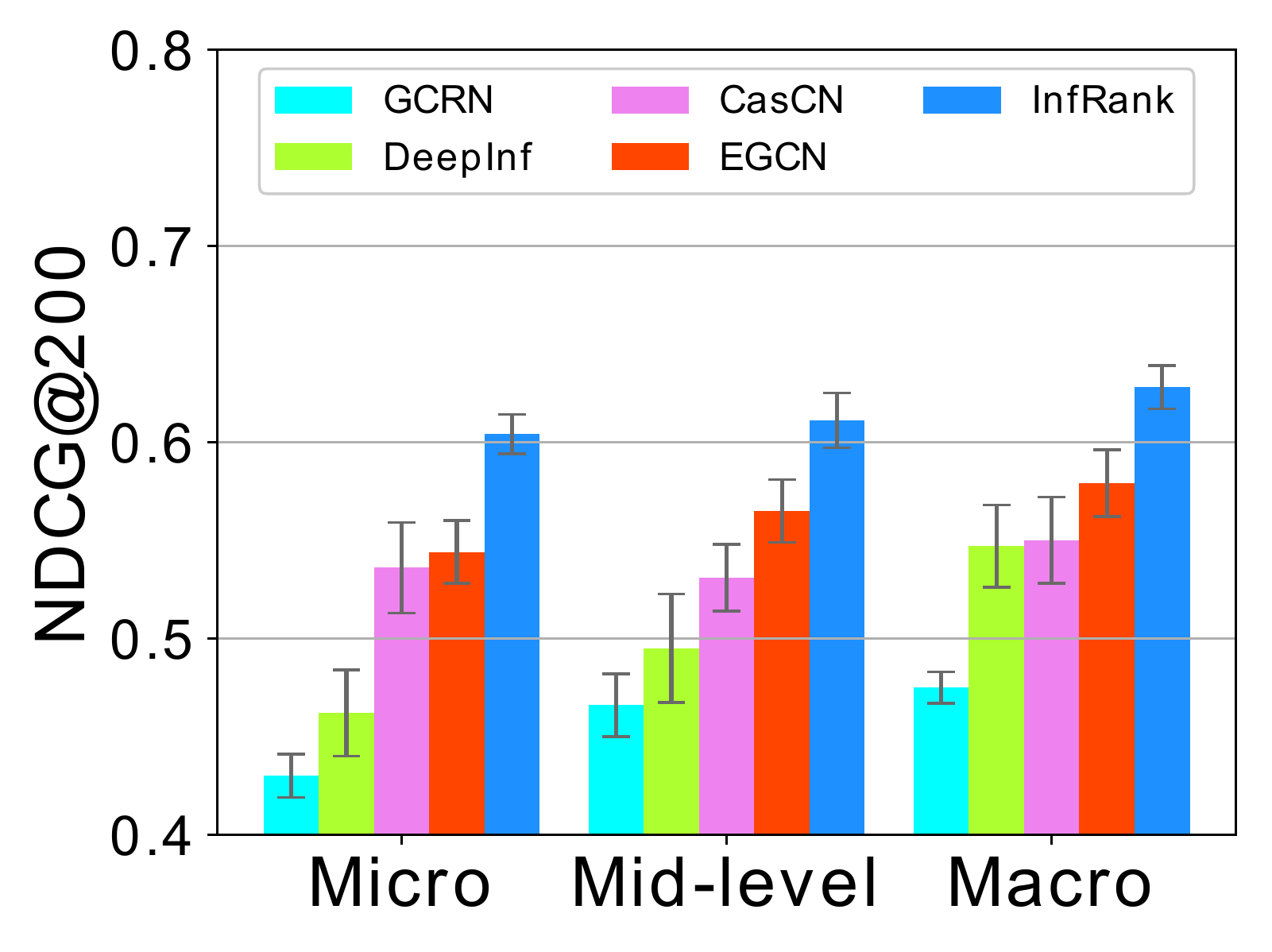}}
\caption{Performance evaluation on influencers with different sizes of followers. \textit{InfluencerRank} shows consistently good performance regardless of the audience size of influencers.}
\label{fig:micro}
\end{figure}

Figure~\ref{fig:micro} shows the average NDCG scores of \textit{InfluencerRank} and four baseline methods, including GCRN~\cite{seo2018structured}, DeepInf~\cite{qiu2018deepinf}, CasCN~\cite{chen2019information}, and EGCN~\cite{pareja2020evolvegcn} over the micro, mid-level, and macro-influencers. The results show that the proposed model has robust performance to discover effective influencers in the groups of all ranges of followers compared to the baseline methods.
More specifically, DeepInf~\cite{qiu2018deepinf} fails to discover effective micro-influencers. 
This is probably because DeepInf disregards the temporal information which is critical to find micro-influencers who have relatively large variance on their features and engagement rates over time compared to macro-influencers who are matured.
On the other hand, our proposed model can accurately find highly effective micro-influencers since their unique features are captured by sequential learning of temporal information.

\section{Conclusion}\label{sec:conclusion}

In this paper, we propose a ranking model to discover influencers with high engagement rates by learning temporal dynamics of their posting behaviors.
To represent the characteristics of influencers and their posting behaviors at each time period, we build a heterogeneous network that consists of influencers and social media elements as nodes, such as hashtags, user tags, and image objects. 
Moreover, each node can be associated with context features in six categories, including node type, profile, image, text, posting, and reaction features.
Based on the GCN-encoded representations of influencers at each timestamp, our proposed model applies attentive RNNs to model historical behaviors of influencers, thereby accurately ranking influencers by their engagement rate scores.
The results of the extensive experiments show that InfluencerRank outperforms existing baseline methods.

\subsection{Broader Impact and Ethical Considerations}
The utility of our proposed framework is expected to significantly increase given a decision of Instagram, one of the most popular influencer marketing platform, that considers to hide the number of likes on each post~\cite{blog2021giving,loren2019hiding} to help mental health issues of social media users~\cite{rsph2014mental}. 
Unlike prior work, the number of likes is not used in discovering influencers in \textit{InfluencerRank}, hence our proposed model can be particularly used by brands with relatively small business sizes, who may be suffering from the heavy expense of discovering effective influencers among millions of candidates~\cite{blog2017advertisers} in a situation where the number of likes is hidden from other users.
Additionally, our model is also capable of adopting additional node features and node types in the network for further improvements. As a result, we believe our model can be widely exploited in finding highly effective influencers for businesses from small retailers to global brands.

{\small
\bibliography{aaai23}
}

\end{document}